# Enhanced superconductivity by Na doping in SnAs-based layered compound $Na_{1+x}Sn_{2-x}As_2$

Hao Yuwen[1,#], Yosuke Goto[1*] Rajveer Jha[1], Akira Miura[2], Chikako Moriyoshi[3], Yoshihiro Kuroiwa[3], Tatsuma D. Matsuda,[1] Yuji Aoki,[1] and Yoshikazu Mizuguchi[1]

[1]*Department of Physics, Tokyo Metropolitan University, Hachioji 192-0397, Japan*

[2]*Faculty of Engineering, Hokkaido University, Kita-13, Nishi-8, Kita-ku, Sapporo, Hokkaido 060-8628, Japan*

[3]*Department of Physical Science, Hiroshima University, 1-3-1 Kagamiyama, Higashihiroshima, Hiroshima 739-8526, Japan*

[#]*Present address: Department of Materials Science and Engineering, Tokyo Institute of Technology, 4259 Nagatsuta, Midori-ku, Yokohama 226-8503, Japan*

E-mail: y_goto@tmu.ac.jp

Superconducting transition temperature ($T_c$) reported in SnAs-based layered compound $NaSn_2As_2$ varies from 1.2 to 1.6 K, implying that its superconductivity is critically sensitive to non-stoichiometry. Here, we demonstrate that Na-doping on the Sn site ($Na_{1+x}Sn_{2-x}As_2$) is effective in enhancing superconductivity, leading to $T_c$ = 2.1 K for $x$ = 0.4. First-principles calculation indicates that such a doping, or $Na_{Sn}$ antisite defects, is energetically favored over other cation vacancies. Our results pave the way for increasing $T_c$ of layered tin pnictide superconductors.

## 1. Introduction

Recently, the reported superconducting transition in the SnAs-based layered compound $NaSn_2As_2$ with a transition temperature ($T_c$) of 1.2–1.6 K has demonstrated that this family of compounds is a new class of van der Waals (vdW)-type layered superconductors[1–3]. These compounds have also been gaining interest for application as thermoelectric[4,5] and topological[6–9] materials. The crystal structure of $NaSn_2As_2$ consists of a buckled honeycomb network of SnAs that is bound by vdW forces and separated by Na ions, as schematically shown in Fig. 1. The structure somewhat resembles that of intercalated graphite[10], HfNCl[11], and transition-metal dichalcogenides (TMDs)[12–15]. Because of the vdW gap between the SnAs layers, $NaSn_2As_2$ can be readily exfoliated through both mechanical and liquid-phase methods, providing a few nanometer-thick single crystals[16,17].

Measurements of the temperature-dependent magnetic penetration depth and thermal conductivity of $NaSn_2As_2$ provide information about the superconducting pairing mechanism of the material and indicate that the superconducting state can be classified into a fully gapped $s$-wave state with atomic-scale disorder[2,3]. The variation in the reported $T_c$ for $NaSn_2As_2$, from 1.2 to 1.6 K, implies that its superconductivity is critically sensitive to non-stoichiometry[1–3]. We reported the synthesis and superconductivity of $Na_{1-x}Sn_2P_2$ with $T_c$ = 2.0 K[18]. Its crystal structure belongs to the $R\bar{3}m$ space group, which is similar to that of $NaSn_2As_2$. Because the band structure and the Debye temperature of these compounds are almost comparable, the higher $T_c$ for $Na_{1-x}Sn_2P_2$ is expected to be due to self-doping by Na vacancy defects.

In this study, we demonstrate the effect of non-stoichiometry on the superconductivity of $NaSn_2As_2$. We report that Na doping on the Sn sites ($Na_{1+x}Sn_{2-x}As_2$) rather than Na vacancies is effective in enhancing superconductivity, which seems to be different from the trend observed in $Na_{1-x}Sn_2P_2$. First-principles calculations indicate that such doping, or $Na_{Sn}$ antisite defects, where $\alpha_\beta$ denotes an antisite of atom $\alpha$ on site $\beta$ [19], is energetically favored over other cation vacancies.

## 2. Experimental Methods

Polycrystalline samples of $Na_{1+x}Sn_{2-x}As_2$ ($x$ = 0, 0.1, 0.2, 0.3, 0.4) were prepared by solid-state reactions using NaAs, Sn (Kojundo Chemical, 99.99%), and As (Kojundo Chemical, 99.9999%) as starting materials. To obtain NaAs, Na (Sigma-Aldrich, 99.9%) and As (1:1 ratio) were heated at 400 °C for 10 h in an evacuated quartz tube. The surface oxide layer on Na was mechanically cleaved before experiments. A mixture of the starting materials (NaAs, Sn, and As) was pressed into a pellet and heated in an evacuated quartz tube for 20 h at 500 °C for $x \leq 0.2$ and at 550 °C for $x \geq 0.3$. The sample preparation procedures were conducted in an Ar-filled glovebox with a gas-purification system. To examine Na vacancy and Sn vacancy defects, $Na_{0.8}Sn_2As_2$ and $NaSn_{1.8}As_2$ were also prepared in the same manner.

Powder X-ray diffraction (PXRD) was measured using CuK$\alpha$ radiation (Rigaku, Miniflex 600 equipped with a D/tex detector). Synchrotron powder X-ray diffraction (SPXRD) of $x$ = 0.3 was measured at BL02B2 of SPring-8 under proposal number 2018B1246. The diffraction data were collected using a high-resolution one-dimensional semiconductor detector, multiple MYTHEN system[20)]. The wavelength of the radiation beam was determined to be 0.496345(1) Å using a $CeO_2$ standard. The crystal structure parameters were refined with the Rietveld method using RIETAN-FP software[21)]. The crystal structure was visualized using VESTA software[22)].

The temperature ($T$) dependence of the electrical resistivity ($\rho$) was measured using the four-probe method; $\rho$–$T$ curves under a magnetic field were measured with a physical property measurement system (PPMS; Quantum Design) equipped with a $^3$He-probe system. Magnetization as a function of $T$ was measured using a superconducting quantum interference device (SQUID) magnetometer (Quantum Design MPMS-3) with an applied field of 10 Oe after both zero-field cooling (ZFC) and field cooling (FC).

Electronic structure calculations based on density functional theory were performed using the VASP code[23,24)]. The exchange-correlation potential was treated within the generalized gradient approximation using the Perdew−Becke−Ernzerhof method[25)]. The Brillouin zone was sampled using a gamma-centered k-point grid of 9 × 9 × 3, and a cutoff of 500 eV was chosen for the plane-wave basis set. For calculations of the defect formation energy, lattice parameters were fixed to those for the undoped composition, and atomic coordinates were optimized until Hellmann–Feynman forces were reduced to less than 0.05 eV Å$^{-1}$.

## 3. Results and Discussion

PXRD patterns of the obtained samples are presented in Fig. 2(a). Almost all the diffraction peaks can be assigned to the trigonal $R\bar{3}m$ space group, indicating that the obtained sample is mainly composed of $Na_{1+x}Sn_{2-x}As_2$. A diffraction peak due to unknown impurities is found in the $x = 0.4$ sample, as denoted by the asterisk. The lattice parameters obtained using Rietveld refinement are shown in Fig. 2(b). It is evident that $x \geq 0.1$ samples show smaller $a$- and larger $c$-axis length than $x = 0$; however, a systematic change with $x$ is not observed.

The $x = 0.3$ sample was used as a representative sample for detailed crystal structure analysis using SPXRD. In addition to the diffraction peaks assigned to $Na_{1+x}Sn_{2-x}As_2$, several peaks attributable to the impurity phases NaOH (12.2 wt%), $Sn_4As_3$ (5.4 wt%), and $Na_2Sn(OH)_6$ (5.0 wt%) are also observed, as shown in Fig. 3. It is believed that hydroxide impurities form during the sample handling of Na-rich samples in an ambient atmosphere. The crystal structure of $NaSn_2As_2$ includes three distinct crystallographic sites: Na, 3*a*; Sn, 6*c*; and As, 6*c*. First, the crystal structure was refined without cation mixed occupation, resulting in a reliability factor $R_{wp}$ of 8.49%. Then, Na/Sn mixed occupation on the 3*a* site was allowed with two constraints: atomic displacement parameters were set equal for Na and Sn in the same position, and the total occupancy was fixed at 100%. The refinement indicates partial occupancies of 97% Na and 3% Sn at the 3*a* position, and $R_{wp}$ was evaluated to be 8.31%. Finally, Na/Sn mixed occupation was allowed on the 6*c* (Sn) site in the same manner, resulting in 96% Sn and 4% Na in this position. $R_{wp}$ was reduced to 8.27% in the final refinement cycle. Therefore, Rietveld fitting results yield the chemical composition of $Na_{1.05}Sn_{1.95}As_2$.

Figures 4(a) and (b) show the temperature-dependent electrical resistivity. Metallic behavior is observed at temperatures above approximately 10 K. The absolute value of the electrical resistivity varies from sample to sample, most likely due to the presence of insulating impurities. It should be noted that no anomaly was observed in the normal conducting region, which is unlike single-crystal $NaSn_2As_2$ prepared using a Sn flux method that exhibits a charge-density-wave-like anomaly at 193 K[2]. Recent measurements of extended x-ray absorption fine structure sfor polycrystalline $NaSn_2As_2$ also detect the anomaly in the interlayer correlation[26].

The samples exhibit zero resistivity at low temperature, indicating a transition to superconducting states. For $x \geq 0.1$, zero resistivity is typically found at temperatures below 1.9 K. Notably, the electrical resistivity of $x = 0.3$ and 0.4 starts to decrease well above 2 K. However, the resistivity drop is relatively wide, which is probably caused by inhomogeneity in these samples. In the present study, we determined $T_c$ to be the temperature at which $\rho$ is 50% of the value at 3 K ($T_c^{50\%}$). Figure 5 depicts the relationship of $T_c^{50\%}$ versus $x$ in the $Na_{1+x}Sn_{2-x}As_2$ sample. $T_c^{50\%}$ in the $x = 0$ sample is 0.9 K, which is lower than that of single crystals reported so far (1.2−1.6 K)[1–3]. $T_c^{50\%}$ tends to increase with increasing $x$, leading to $T_c^{50\%} = 2.1$ K for the $x = 0.4$ sample.

The temperature-dependent magnetization is presented in Fig. 6. For $x \geq 0.1$, diamagnetic signals corresponding to the superconducting transition are seen below 2 K, which is consistent with the resistivity measurements. The $x = 0.3$ sample exhibits clear diamagnetic signals, indicating the bulk nature of the superconductivity, whereas diamagnetization in the other samples is noticeably weak in the measured temperature region, i.e., above 1.8 K. We deduce that weak diamagnetic signals in the $x = 0.1$ and 0.2 samples are due to lower $T_c$ than in the $x = 0.3$ sample. At the same time, the weak diamagnetic signal of $x = 0.4$ may arise from the presence of unknown impurity phases, as mentioned earlier. Magnetization measurements below 1.8 K will be required to examine the bulk nature of superconductivity in samples with stoichiometries other than when $x = 0.3$.

The electrical resistivity of the $x = 0.3$ sample under a range of magnetic field strengths is presented in Fig. 7(a). The transition temperature shifts to lower temperatures with increasing applied magnetic field. It is noteworthy that the superconducting transition is distinctly broadened under a magnetic field, which is probably caused by the anisotropic upper critical field that is generated by the two-dimensional layered crystal structure. Figure 7(b) presents the temperature versus magnetic field phase diagram. On the basis of the Werthamer–Helfand–Hohemberg (WHH) model in the dirty limit[27], the upper critical field at 0 K is estimated to be $\mu_0 H_{c2}(0)$ = ca. 2.6 T. The coherence length $\xi$ is estimated to be approximately 11 nm using the equation $\xi^2 = \Phi_0/2\pi\mu_0 H_{c2}$, where $\Phi_0$ is the magnetic flux quantum.

We compare the superconductivity of $Na_{1+x}Sn_{2-x}As_2$ with that of $Na_{0.8}Sn_2As_2$ and $NaSn_{1.8}As_2$ to examine the influence of the Na vacancy and Sn vacancy defects. As shown in Fig. 8, the PXRD patterns of these samples show the presence of additional impurity phases, specifically SnAs in $Na_{0.8}Sn_2As_2$ and $Sn_4As_3$ in $NaSn_{1.8}As_2$. The lattice parameters are plotted in Fig. 2(b). Figure 9(a) shows the $\rho$–$T$ curves for $Na_{0.8}Sn_2As_2$; the two-step resistivity drops at 1.0 and 3.0 K are most likely due to the superconducting transition of $Na_{0.8}Sn_2As_2$ and SnAs, respectively. Because SnAs is known to be a type-I superconductor with a critical field of less than 20 mT[28)], superconductivity due to the SnAs secondary phase is easily suppressed by applying a magnetic field, as shown in Fig. 9(a). On the contrary, zero resistivity is not obtained down to 1.5 K for $NaSn_{1.8}As_2$, although the onset $T_c$ was evaluated to be 1.7 K (Fig. 9(b)). These results confirm that Na doping on the Sn sites ($Na_{1+x}Sn_{2-x}As_2$) is more effective for emerging superconductivity than Na vacancy and Sn vacancy defects.

Finally, we briefly discuss the defect formation in $NaSn_2As_2$ by using first-principles calculations. Na doping on the Sn sites ($Na_{Sn}$ antisite defects), Na vacancies, and Sn vacancies were considered as possible sources of defect formation. The formation energy for $Na_{Sn}$ antisite defects was calculated to be 0.86 eV and that of Na vacancies and Sn vacancies was evaluated to be 1.70 and 1.67 eV, respectively, which is summarized in Table I. Therefore, Na doping on Sn sites is energetically favorable in comparison to the formation of Na or Sn vacancies, which is consistent with the experimental results.

We also investigated the defect formation of $NaSn_2P_2$ to discuss the difference between $NaSn_2As_2$ and $Na_{1-x}Sn_2P_2$. As summarized in Table I, the calculated results for $NaSn_2P_2$ are qualitatively similar to those for $NaSn_2As_2$. The formation energy of $Na_{Sn}$ antisite defects is lower than that of Na and Sn vacancies. This result seems to contradict the experimental work on $Na_{1-x}Sn_2P_2$, where Na vacancies were indicated in the crystal structure analysis by synchrotron XRD[18)]. However, it is known that $E_{form}$ depends on the Fermi energy and the charge of defects[19,29–31)]. Further study will be required to elucidate the detailed defect physics/chemistry of these compounds.

Let us stress that we focus on hole doping for emerging superconductivity in SnAs-based layered $NaSn_2As_2$. For example, $EuSn_2As_2$[17)] or $Li_{1-x}Sn_{2+x}As_2$[4)], which are considered to be electron-doped systems of $NaSn_2As_2$, are not superconductive until 2 K. $SrSn_2As_2$ has been reported to be in the vicinity of a quantum critical point[6,7)], but our preliminary experiments show that it is also non-superconducting down to 0.5 K (data are not shown). In other words, these compounds might also be turned into superconductors via hole doping. Our experimental results on the effects of impurity doping in these compounds will be published in the near future.

## 4. Conclusions

In summary, we report non-stoichiometry and superconductivity in the SnAs-based layered compound $NaSn_2As_2$. It is demonstrated that Na doping on Sn sites is effective in increasing $T_c$, leading to a $T_c$ of 2.1 K for $x = 0.4$ in $Na_{1+x}Sn_{2-x}As_2$. This is in contrast to $Na_{1-x}Sn_2P_2$ that contains Na vacancy defects. First-principles calculations indicate that such doping, or $Na_{Sn}$ antisite defects, is energetically favored over cation vacancies. Our results pave the way for increasing $T_c$ beyond 2 K in layered tin pnictide superconductors.

## Acknowledgments

We thank O. Miura for use of MPMS; R. Higashinaka for use of PPMS. This work was partly supported by Grants-in-Aid for Scientific Research (Nos. 15H05886, 15H05884, 16H04493, 17K19058, 15H03693, and 19K15291) and Iketani Science and Technology Foundation (No. 0301042-A), Japan.

## Figure Captions

**Fig. 1.** Crystallographic structure of $Na_{1+x}Sn_{2-x}As_2$ ($x$ = 0.3). Na and Sn sites show slightly mixed occupation (see text for details).

**Fig. 2.** (a) Powder X-ray diffraction (PXRD) patterns for $Na_{1+x}Sn_{2-x}As_2$. Vertical tick marks at the bottom indicate the calculated Bragg diffraction positions of $NaSn_2As_2$. The asterisk denotes an unknown impurity phase. (b) Lattice parameters as a function of $x$. For $x$ = 0.3, both $a$ and $c$ calculated using SPXRD almost coincide with those obtained from the experimental PXRD data. The lattice parameters of $Na_{0.8}Sn_2As_2$ and $NaSn_{1.8}As_2$ are also shown.

**Fig. 3.** SPXRD pattern ($\lambda$ = 0.496345(1) Å) and the results of Rietveld refinement for $x$ = 0.3. The red circles and solid curve represent the observed and calculated patterns, respectively, and the difference between the two is shown in blue. Vertical tick marks indicate the calculated Bragg diffraction positions for $Na_{1+x}Sn_{2-x}As_2$, NaOH, $Sn_4As_3$, and $Na_2Sn(OH)_6$, from top to bottom. The reliability factors are $R_{wp}$ = 8.27%, $R_B$ = 3.62%, and GOF = 5.79.

**Fig. 4.** (a) Temperature ($T$) dependence of the electrical resistivity ($\rho$) of $Na_{1+x}Sn_{2-x}As_2$. (b) $\rho$–$T$ curves below 6 K. The $\rho$–$T$ data for $x$ = 0 were multiplied by 10 for clarity.

**Fig. 5.** $T_c^{50\%}$ versus $x$ of polycrystalline $Na_{1+x}Sn_{2-x}As_2$ (red closed circle). For comparison, $T_c^{50\%}$ of single-crystal $Na_{1+x}Sn_{2-x}As_2$ where $x$ = 0 was taken from the literature (black open squares)[1,3].

**Fig. 6.** Temperature ($T$) dependence of the magnetization ($M$) of $Na_{1+x}Sn_{2-x}As_2$ from 1.8 to 5 K. The magnetization measured after zero field cooling (ZFC) and field cooling (FC) almost coincides except when $x$ = 0.3.

**Fig. 7.** (a) Resistivity ($\rho$) versus temperature ($T$) curves for $x$ = 0.3 under magnetic fields ($\mu_0 H$). (b) Magnetic field–temperature phase diagram for $x$ = 0.3. The solid line was calculated using the WHH model.

**Fig. 8.** Powder X-ray diffraction (PXRD) patterns for $Na_{0.8}Sn_2As_2$ and $NaSn_{1.8}As_2$. Vertical tick marks at the bottom represent the calculated Bragg diffraction positions for $NaSn_{1.8}As_2$. Diffraction peaks due to impurity phases (SnAs and $Sn_4As_3$) are denoted by arrows.

**Fig. 9.** Resistivity ($\rho$) versus temperature ($T$) curves for (a) $Na_{0.8}Sn_2As_2$ and (b) $NaSn_{1.8}As_2$. A resistivity drop at 3.0 K in (a) is most likely due to a superconducting transition of the SnAs secondary phase.

**Table I**

Chemical reactions and defect formation energy ($E_{form}$) of $NaSn_2As_2$ and $NaSn_2P_2$

| Parent compound | Chemical reaction | $E_{form}$ (eV) | Note |
|---|---|---|---|
| $NaSn_2As_2$ | $(NaSn_2As_2)_3$ + Na → $(Na_{1.33}Sn_{1.67}As_2)_3$ + Sn | 0.87 | $Na_{Sn}$ antisites |
| | $(NaSn_2As_2)_3$ → $(Na_{0.67}Sn_2As_2)_3$ + Na | 1.70 | Na vacancies |
| | $(NaSn_2As_2)_3$ → $(NaSn_{1.67}As_2)_3$ + Sn | 1.67 | Sn vacancies |
| $NaSn_2P_2$ | $(NaSn_2P_2)_3$ + Na → $(Na_{1.33}Sn_{1.67}P_2)_3$ + Sn | 1.29 | $Na_{Sn}$ antisites |
| | $(NaSn_2P_2)_3$ → $(Na_{0.67}Sn_2P_2)_3$ + Na | 1.97 | Na vacancies |
| | $(NaSn_2P_2)_3$ → $(NaSn_{1.67}P_2)_3$ + Sn | 2.14 | Sn vacancies |

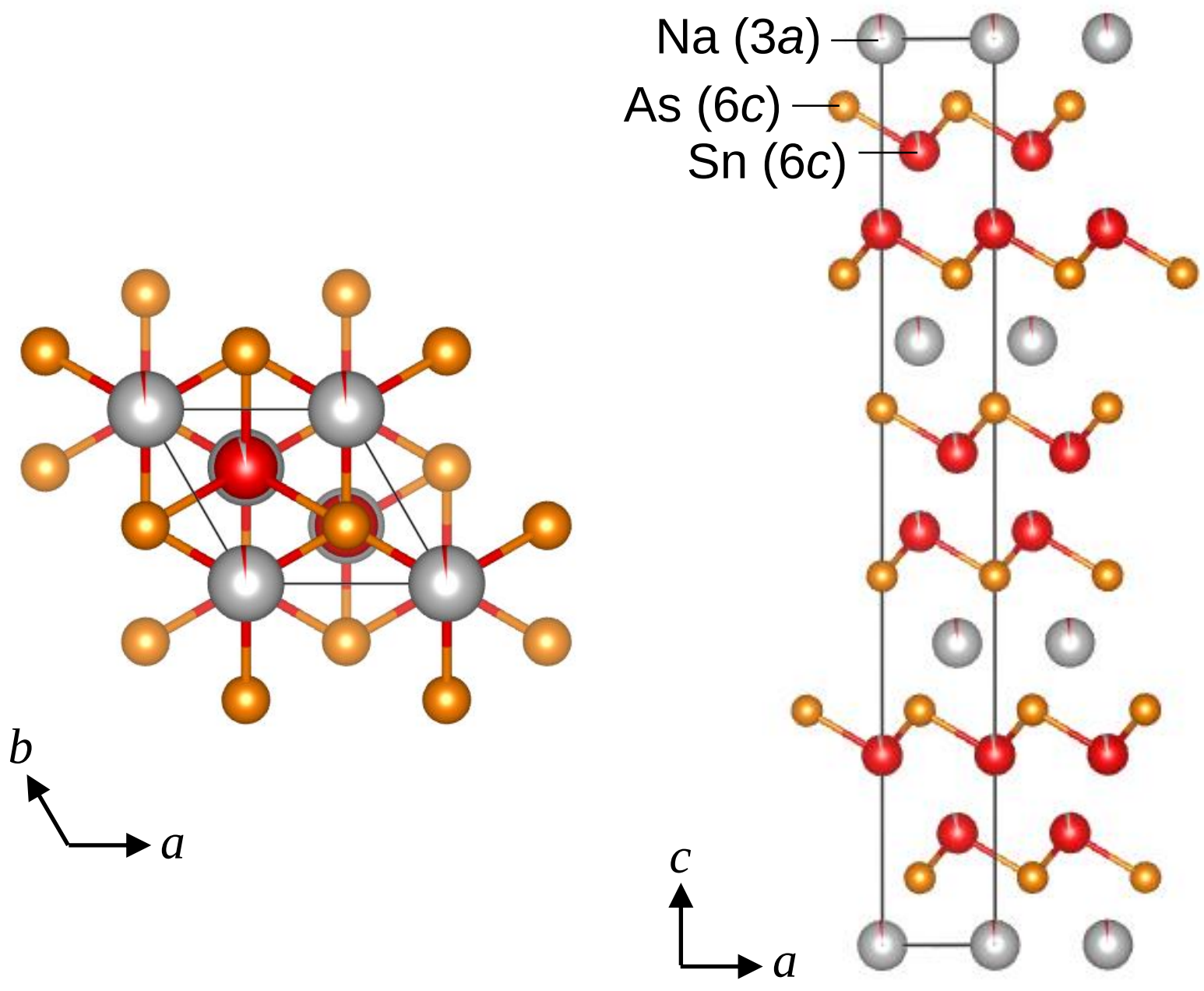

**Fig. 1.**

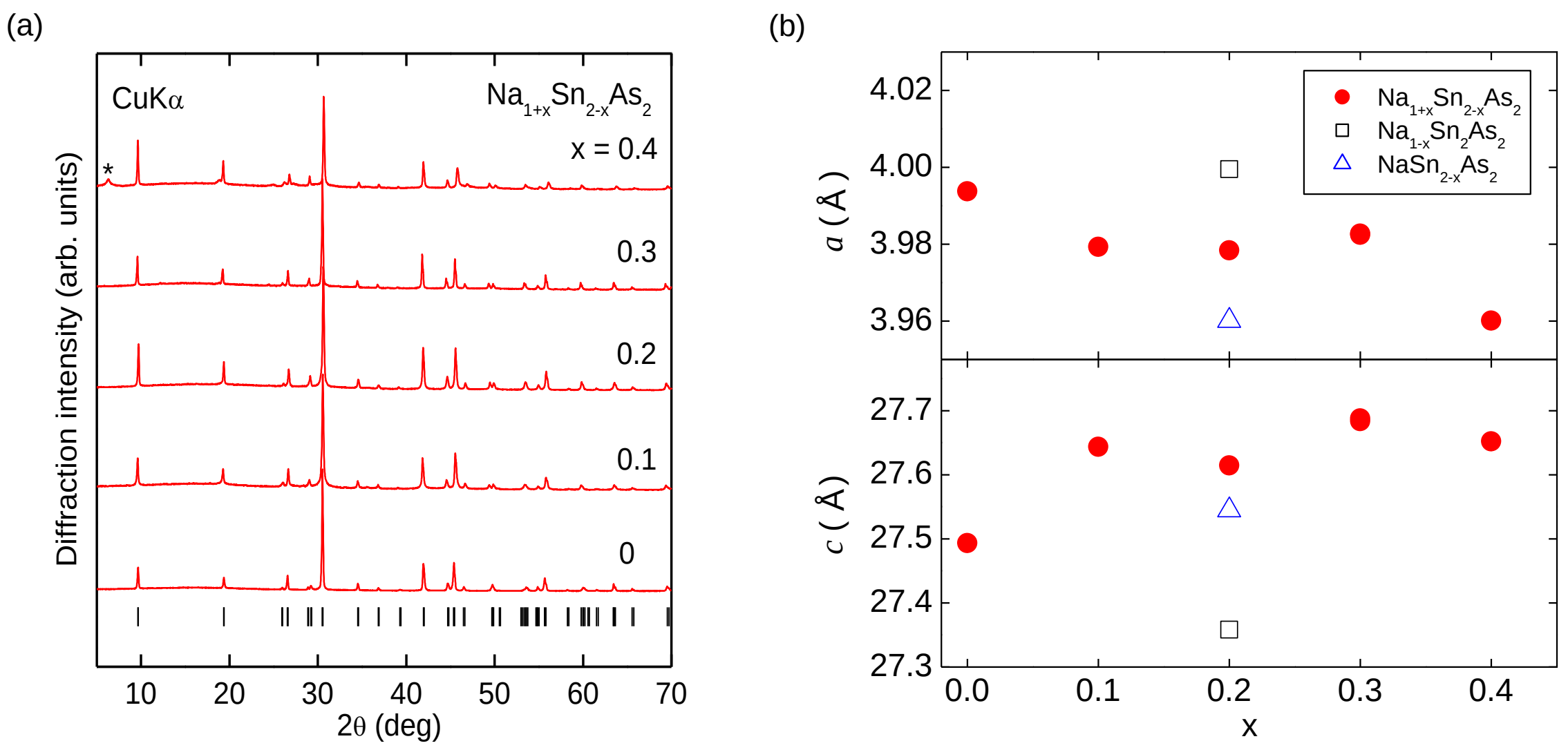

**Fig. 2.**

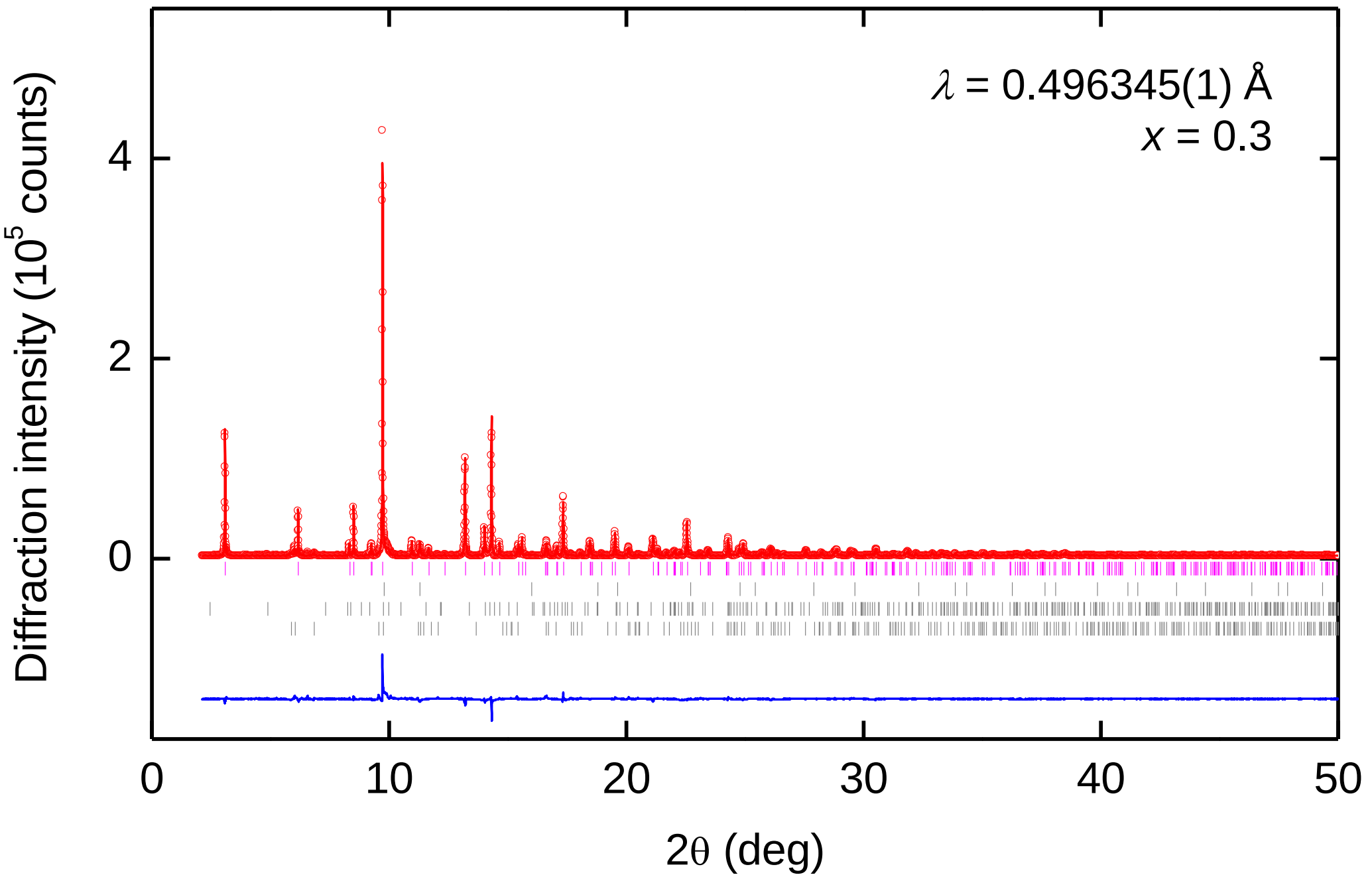

**Fig. 3.**

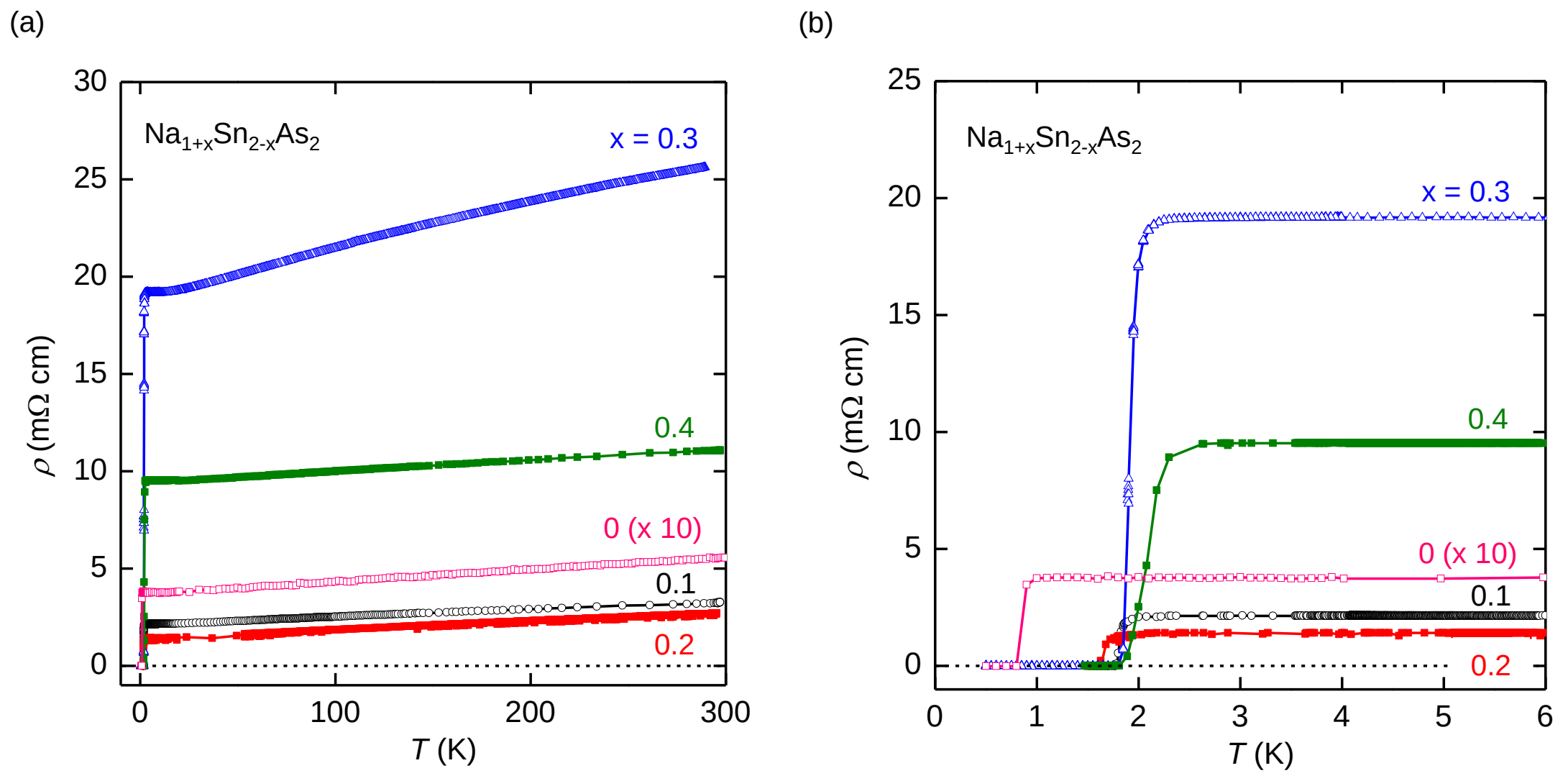

**Fig. 4.**

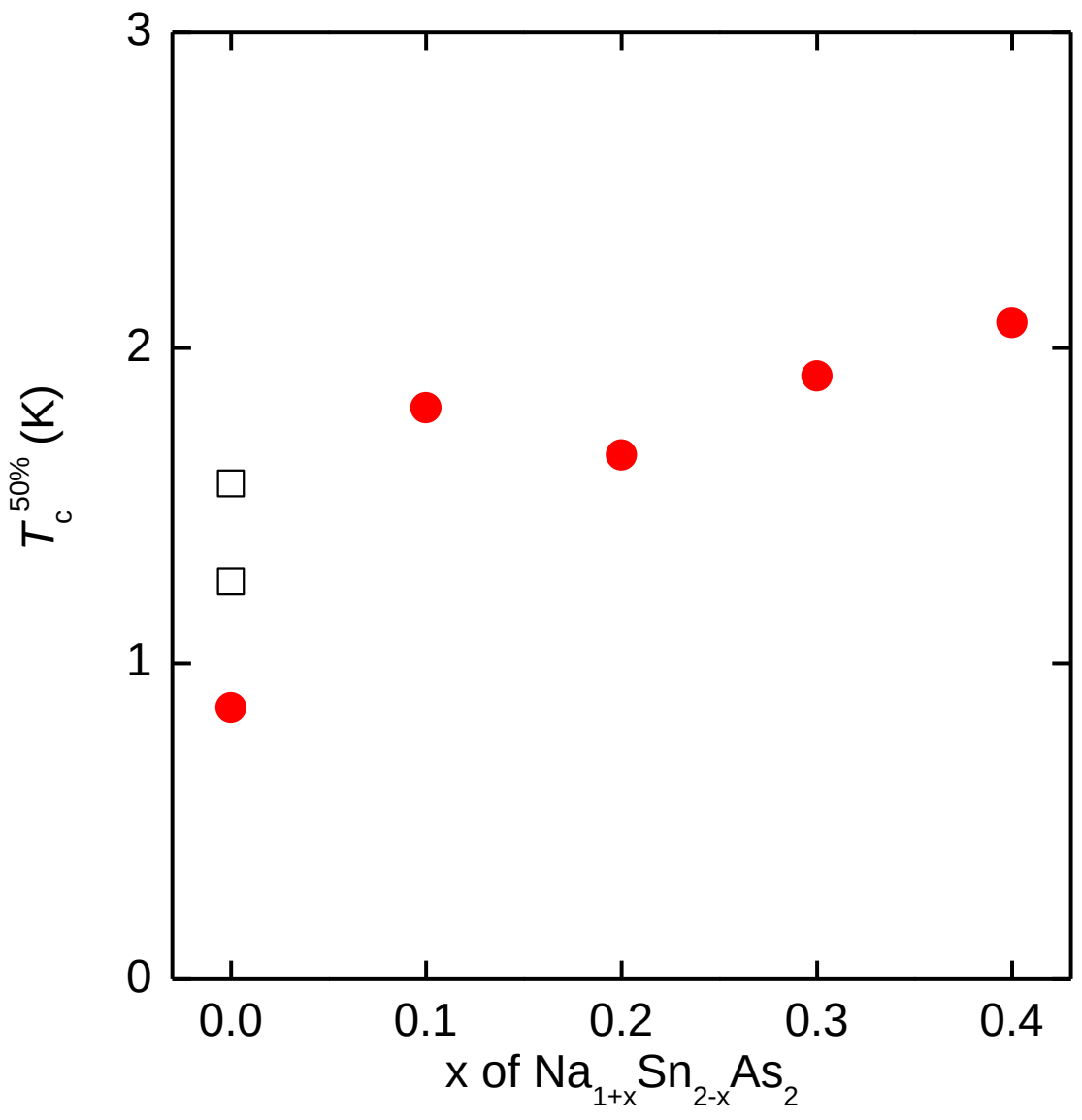

**Fig. 5.**

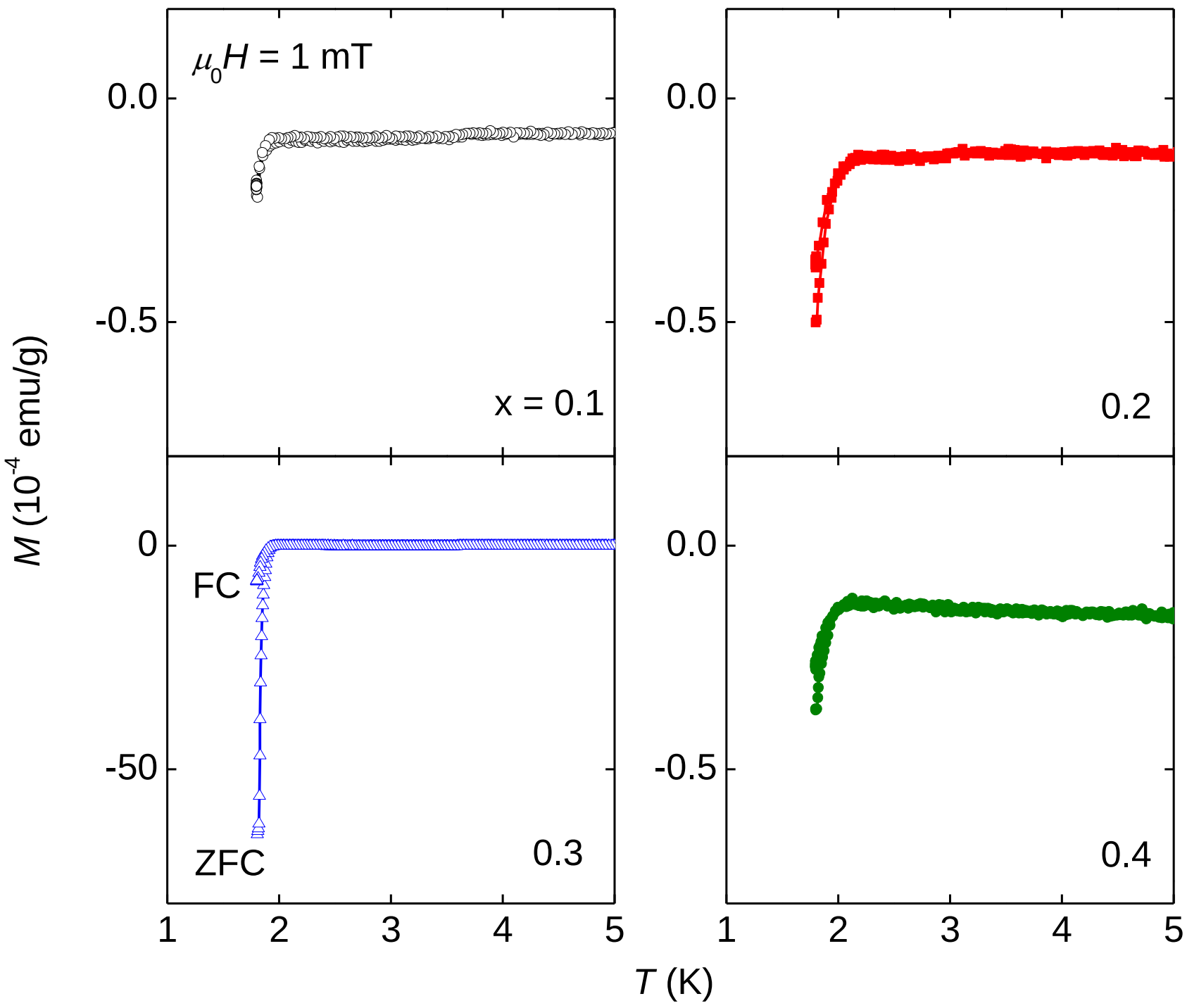

**Fig. 6.**

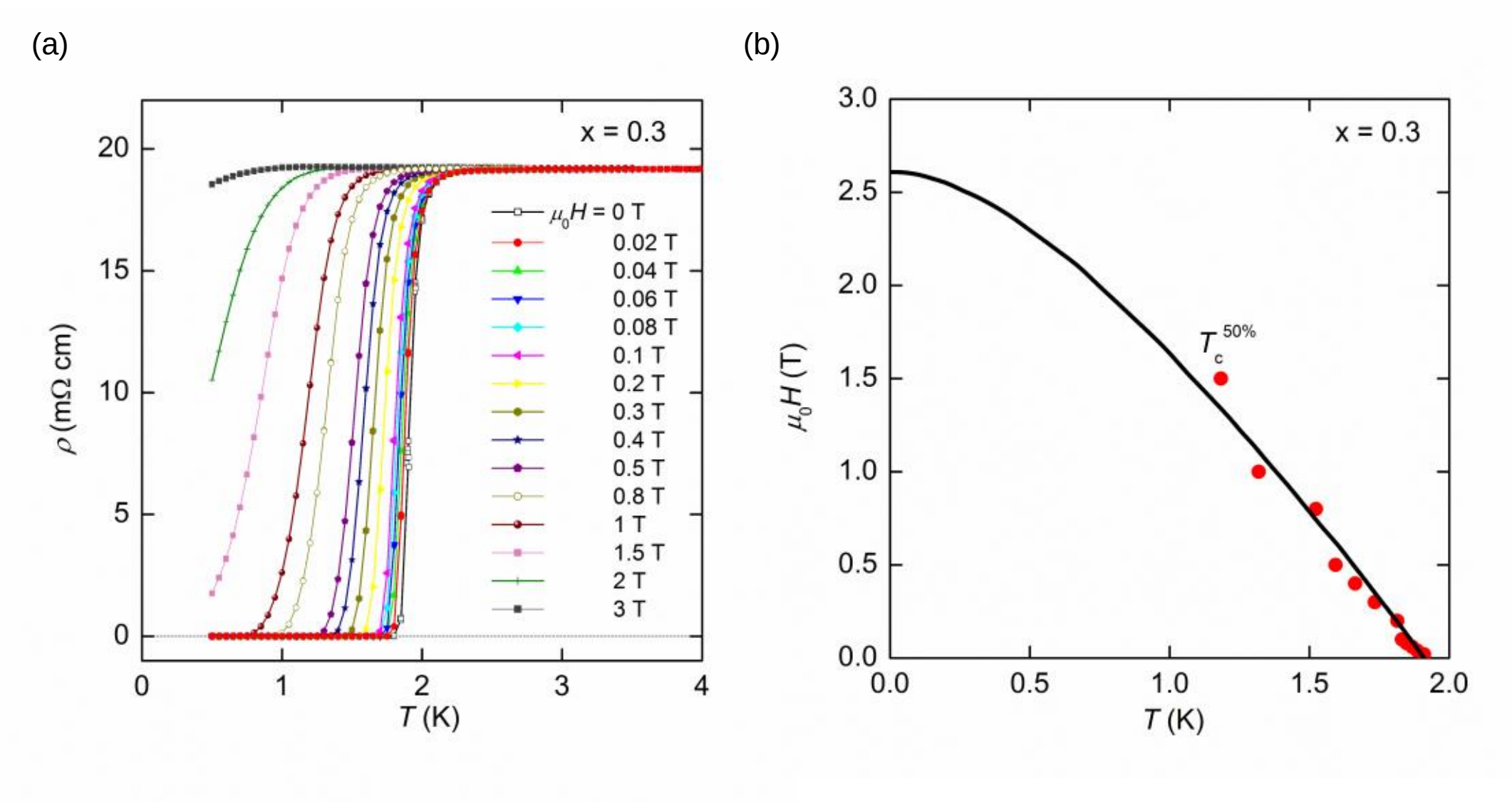

**Fig. 7.**

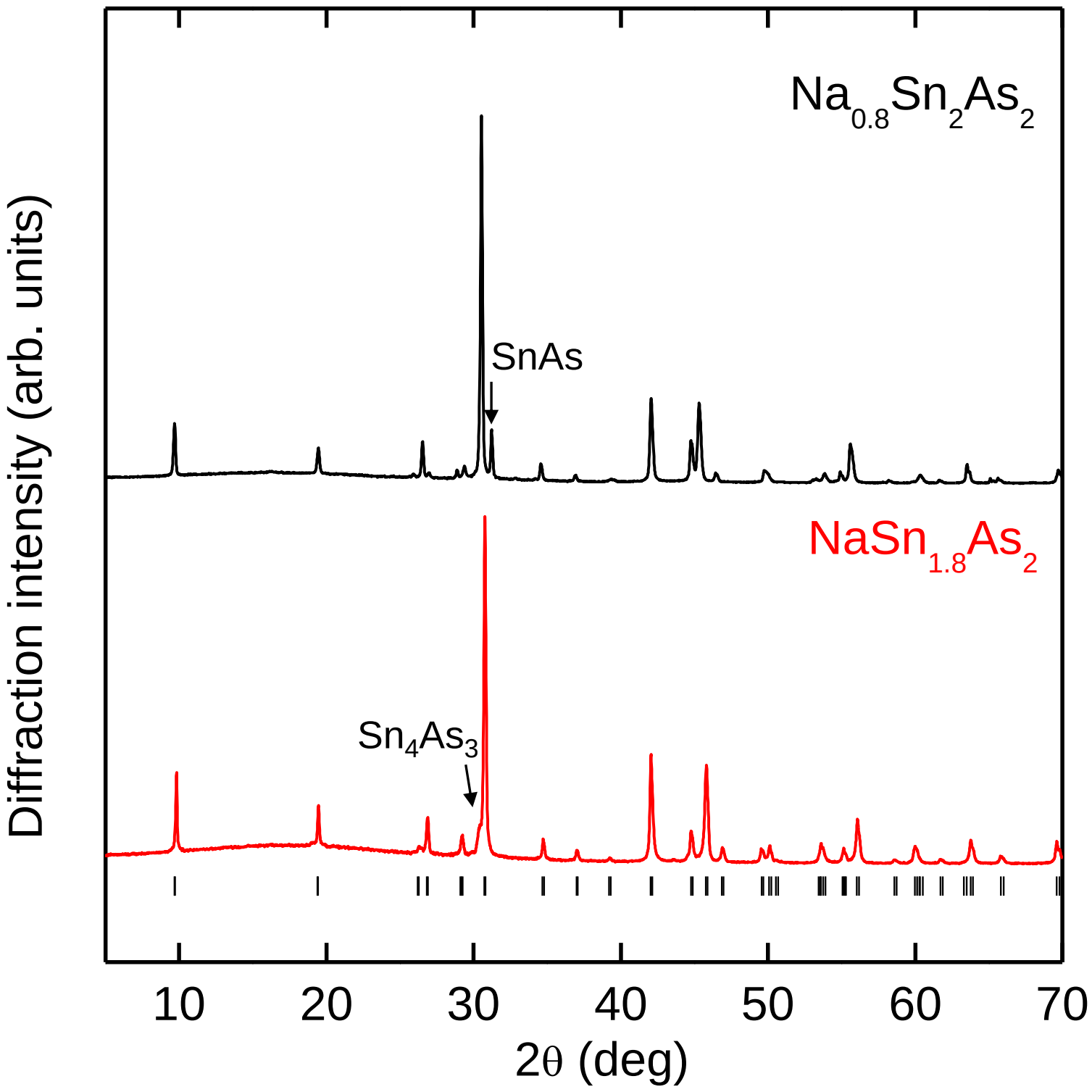

**Fig. 8.**

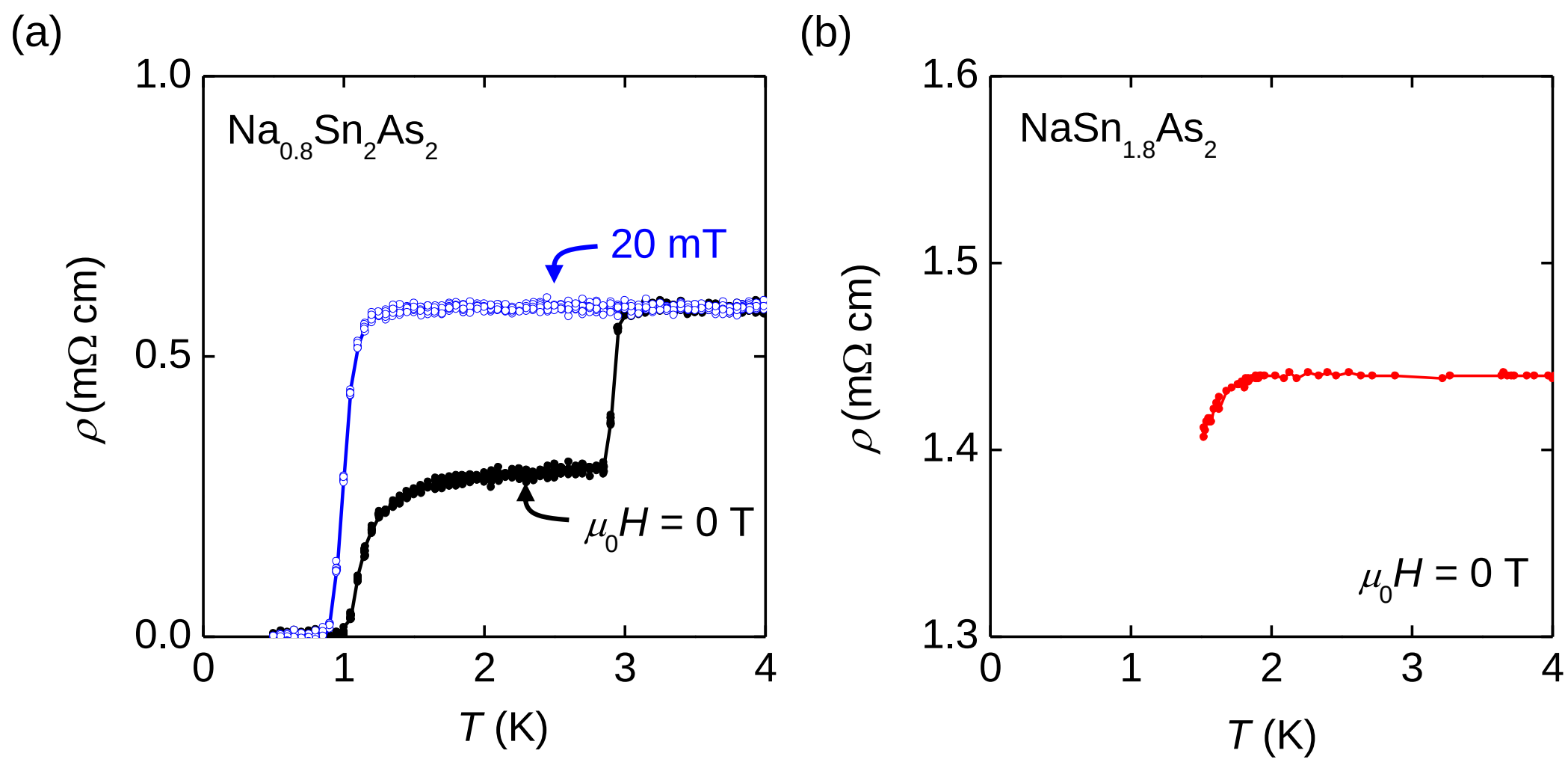

**Fig. 9.**